\documentclass[12pt]{article}
\usepackage{amssymb}
\usepackage{amsthm,amsmath}
\usepackage{accents}
\usepackage{colordvi}
\usepackage{graphicx}
\usepackage{epstopdf}

\newcommand{\onetom}{1,\cdots,m}

\graphicspath{{figs/}}

\begin{document}
\begin{center}
{\large\bf  On complex networks with pinning controllers }

\footnote{This work is jointly supported by the National Key R$\&$D Program of China (No.
2018AAA010030), National Natural Sciences Foundation of China under Grant (No.
62072111).
}
\begin{center}
Tianping Chen\footnote{Tianping Chen is with the School of Mathematics, Fudan University, 200433, Shanghai, China.}

\end{center}
\end{center}

{\bf Abstract}
In this short paper, we explore relationship between various models of complex networks with pinning controllers.


In the paper [1], we discussed following coupling network with a single pinning controller
\begin{eqnarray}
\left\{\begin{array}{cc}\frac{dx_1(t)}{dt}&=f(x_1(t),t)+c\sum\limits_{j=1}^ma_{1j}x_j(t)\\
&-c\varepsilon(x_{1}(t)-s(t)),\\
\frac{dx_i(t)}{dt}&=f(x_i(t),t)+c\sum\limits_{j=1}^ma_{ij}x_j(t),\\
&i=2,\cdots,m \end{array}\right. 
\end{eqnarray}
where $a_{ij}\ge 0$, if $i\ne j$ and $a_{ii}=-\sum_{j\ne i}a_{ij}$.

It is clear that (1) can be written as
\begin{eqnarray}
\frac{d(x_i(t)-s(t))}{dt}&=f(x_i(t),t)-f(s(t),t)
+c\sum\limits_{j=1}^m\tilde{a}_{ij}(x_j(t)-s(t))
\end{eqnarray}
where
\[\tilde{A}=\left(\begin{array}{cccc}a_{11}-\varepsilon&a_{12}&\cdots&a_{1m}\\
a_{21}&a_{22}&\cdots&a_{2m}\\\vdots&\vdots&\ddots&\vdots\\
a_{m1}&a_{m2}&\cdots&a_{mm}\end{array}\right)\]

In [1], we gave 

{\bf Proposition 1} Let $\Xi=[\xi_{1},\cdots,\xi_{n}]^{T}$ be the left eigenvector of the matrix $A$ corresponding to eigenvalue $0$. Denote $(\Xi \tilde{A})^{s}=\frac{\Xi \tilde{A}+\tilde{A}^{T}\Xi}{2} $. Then, $(\Xi \tilde{A})^{s}$ is negative definite with eigenvalues $0>\lambda_{1}\ge \cdots,\lambda_{m}$.

and proved following Theorems

{\bf Theorem A} Suppose $A$ is symmetric. $0>\lambda_1>\cdots>\lambda_m$ are the eigenvalues
of $\tilde{A}$. If there are  positive diagonal matrices
$P=diag\{p_1,\cdots,p_{n}\}$,
$\Delta=diag\{\Delta_1,\cdots,\Delta_n\}$ and a constant $\eta>0$,
such that QUAD condition
\begin{equation}
(x-y)^{\top}P(f(x,t)-\Delta x-f(y,t)+\Delta y)
\leq -\eta (x-y)^{\top}(x-y)
\end{equation}
is satisfied
and $\Delta_i+c\lambda_1<0$, for $i=1,\cdots,n$. Then, the
controlled system (2) is globally exponentially synchronized to
$s(t)$.

{\bf Theorem B} Suppose $A$ is asymmetric. $0>\lambda_1>\cdots>\lambda_m$ are the eigenvalues
of $\{\Xi\tilde{A}\}^{s}$. If there are  positive diagonal matrices
$P=diag\{p_1,\cdots,p_{n}\}$,
$\Delta=diag\{\Delta_1,\cdots,\Delta_n\}$ and a constant $\eta>0$,
such that QUAD condition
\begin{equation}
(x-y)^{\top}P(f(x,t)-\Delta x-f(y,t)+\Delta y)
\leq -\eta (x-y)^{\top}(x-y)\label{quad}
\end{equation}
is satisfied,
and $\Delta_i+c\lambda_1<0$, for $i=1,\cdots,n$. Then, the
controlled system (2) is globally exponentially synchronized to
$s(t)$.

A few years later, in [2], authors discussed
\begin{eqnarray}
\left\{\begin{array}{cc}\frac{dx_1(t)}{dt}&=f(x_1(t),t)
+c\sum\limits_{j=1}^mG_{1j}\Gamma x_j(t)\\
&-c\varepsilon\Gamma(x_{1}(t)-s(t)),\\
\frac{dx_i(t)}{dt}&=f(x_i(t),t)+c\Gamma\sum\limits_{j=1}^mG_{ij}x_j(t),\\
&i=2,\cdots,m \end{array}\right. 
\end{eqnarray}
where where $G_{ij}\ge 0$, if $i\ne j$ and $G_{ii}=-\sum_{j\ne i}G_{ij}$, $G_{ij}=G_{ji}$. $f(x,t)$ satisfies
\begin{equation}
(x-y)^{\top}(f(x)-f(y))-(x-y)^{\top}K\Gamma(x-y)<0
\end{equation}
where $\Gamma$ is positive definite and $K,\Gamma$ are commutable.

It is equivalent to
\begin{equation}
(x-y)^{\top}(f(x)-f(y))-(x-y)^{\top}K\Gamma(x-y)<-\eta (x-y)^{\top}(x-y)
\end{equation}
for some constant $\eta>0$.

In [3], $G_{ij}$ is assumed as asymmetric.

Now, we will reveal that the results given in [2], [3] are direct consequences of those given in [1].

Firstly, we give following simple proposition on the normal operators in functional analysis.

{\bf Proposition 2}
The assumption that $\Gamma$ and $K$ are commutable is equivalent to that $\Gamma$ and $K$  have same eigenvectors.

In fact, if $\gamma_{i}$ is a eigenvalue of $\Gamma$, $x_{i}$ is the corresponding eigenvector. Then
$$\Gamma x_{i}=\gamma_{i} x_{i}$$
Because $\Gamma K=K\Gamma$, then
$$\Gamma K x_{i}=K\Gamma x_{i}=\gamma_{i} K x_{i}$$
which means $Kx_{i}$ is also an eigenvector of $\Gamma$ corresponding to $\gamma_{i}$~
$$Kx_{i}=k_{i} x_{i}$$
~$\Gamma$ and $K$ have same eigenvectors.

It is well known that the definite matrix  $\Gamma$ can be decomposed as $\Gamma=Q^{T}\bar{\Gamma}Q$, where $Q$ is an orthogonal matrix and $\bar{\Gamma}=diag[\gamma_{1},\cdots,\gamma_{n}]$ is a positive diagonal 

Based on Proposition 2, $K$ can also be written as $K=Q^{T}\bar{K}Q$, $\bar{K}=diag[k_{1},\cdots,k_{n}]$.

Let $\tilde{x}_{i}(t)=Q x_{i}(t)$. $\tilde{s}(t)=Qs(t)$,
and
$$\tilde{f}(u)=Q(f(Q^{-1}u,t)$$
It is clear that
\begin{align}
&(\tilde{x}-\tilde{y})^{\top}(\tilde{f}(\tilde{x})-\tilde{f}(\tilde{y}))
=(x-y)^{T}(f(x)-f(y))
\end{align}

Therefore, in the following, we will discuss following systems
\begin{eqnarray}
\frac{d({x}_i(t)-{s}(t))}{dt}&={f}({x}_i(t))-{f}({s}(t)))
+c\sum\limits_{j=1}^m\tilde{G}_{ij}\bar{\Gamma}({x}_i(t)-{s}(t))
\end{eqnarray}
where $\bar{\Gamma}=diag[\gamma_{1},\cdots,\gamma_{n}]$ is a positive diagonal matrix.

The QUAD condition is
\begin{align}
({x}-{y})^{\top}({f}({x})-{f}(\tilde{y}))
-({x}-{y})^{\top}\bar{K}\bar{\Gamma}({x}-{y})<-\eta ({x}-{y})^{\top}({x}-{y})
\end{align}
where $\bar{K}=diag[k_{1},\cdots,k_{n}]$ is a positive diagonal matrix.

Following the approach proposed in [1], we can prove

{\bf Theorem 1} Suppose $A$ is asymmetric. $0>\lambda_1>\cdots>\lambda_m$ are the eigenvalues
of $\{\Xi\tilde{A}\}^{s}$. If there is a constant $\eta>0$,
such that QUAD condition
\begin{equation}
(x-y)^{\top}[f(x)-f(y)-\bar{\Gamma}\bar{K}(x-y)]
\leq -\eta (x-y)^{\top}(x-y)\label{quad}
\end{equation}
is satisfied and
$\xi_{i}\bar{k}_{j}+c\lambda_{1}<0$ for all $i=1,\cdots,m$, $j=1,\cdots,n$. Then, 
the
controlled system (9)
 is globally exponentially synchronized to
$s(t)$.

{\bf Proof}
Follow [1], define Lyapunov function
\begin{align}
V(\tilde{x}(t))=\frac{1}{2}\sum\limits_{i=1}^m\xi_{i}
({x}_i(t)-{s}(t))^{T}({x}_i(t)-{s}(t))
\end{align}
Differentiating, we have
\begin{align*}
\frac{d(V({x}(t)))}{dt}&=\sum\limits_{i=1}^m\xi_{i}({x}_i(t)-{s}(t))^{T}
[{f}({x}_i(t))-{f}({s}(t))]\\
&+c\sum\limits_{i=1}^m\sum\limits_{j=1}^m\xi_{i}\tilde{G}_{ij}
({x}_i(t)-{s}(t))^{T}\bar{\Gamma}({x}_j(t)-{s}(t))\\
&<\sum\limits_{i=1}^m\xi_{i}({x}_i(t)-{s}(t))^{T}\bar{K}\bar{\Gamma}
({x}_i(t)-{s}(t))\\
&+c\lambda_{1}\sum\limits_{i=1}^m({x}_i(t)-{s}(t))^{T}\bar{\Gamma}
({x}_i(t)-{s}(t))\\&
-\eta \sum\limits_{i=1}^m\xi_{i}({x}_i(t)-{s}(t))^{T}
({x}_i(t)-{s}(t))
\end{align*}
In case that $\xi_{i}\bar{k}_{j}+c\lambda_{1}<0$ for all $i=1,\cdots,m$, $j=1,\cdots,n$,
we have
\begin{align}
\dot{V}({x}(t))< -\eta V({x}(t))
\end{align}
and
\begin{align}
V({x}(t))=O(e^{-\eta t})
\end{align}
which implies
\begin{align}
{x}_{i}(t)-{s}(t)=O(e^{-\eta t}),~~i=1,\cdots,m
\end{align}

{\bf Remark 1} Proposition 1 plays the key role in the theoretical analysis. It concludes that $\Gamma$
and $K$ have same eigenvectors.

{\bf Remark 2} The proof given in [2], [3] is questionable. A key problem is that $K\Gamma=(\frac{K+K^{\top}}{2})\Gamma\le ||\theta||\Gamma$, where
$||\theta||=\lambda_{max}(\frac{K+K^{\top}}{2})$, is incorrect. In general, even though both matrices $A$ and $B$ are positive definite, it is not known whether the product $AB$ is still positive definite. That is one can not prove $x^{T}ABx\le ||A||x^{T}Bx$. Unless $A$ and $B$ have same eigenvectors.

{\bf Remark 3} In [3], the authors proved 

Lemma 2.11. Suppose that $G$ is irreducible and satisfies $\sum_{j=1}^{m}G_{ij}=0$ with $G_{ij}\ge 0$.
Then, there is a positive vector $x$ such that $G^{T}x=0$.

It is clear that this Lemma is a special case of the following Lemma (see Lemma 1 in [4]).

{\bf Lemma 1}\quad 
If $A\in \mathbf A1$, then the following items are valid:
\begin{enumerate}
\item If $\lambda$ is an eigenvalue of $A$ and $\lambda\neq 0$,
then  $Re(\lambda)<0$;

 \item $A$ has an eigenvalue $0$ with
multiplicity 1 and the right eigenvector $[1,1,\dots,1]^{\top}$;

\item  Suppose $\xi=[\xi_{1},\xi_{2},\cdots,\xi_{m}]^{\top}\in
R^{m}$ (without loss of generality, assume
$\sum\limits_{i=1}^{m}\xi_{i}=1$) is the left eigenvector of $A$
corresponding to eigenvalue $0$. Then,  $\xi_{i}\ge 0$ holds for
all $i=\onetom$; more precisely,

\item $A\in\bf A_{2}$ if and only if  $\xi_{i}>0$ holds for all
$i=\onetom$;

\item $A$ is reducible if and only if for some $i$, $\xi_{i}=0$.
In such case, by suitable rearrangement, assume that
$\xi^{\top}=[\xi_{+}^{\top},\xi_{0}^{\top}]$, where
$\xi_{+}=[\xi_{1},\xi_{2},\cdots,\xi_{p}]^{\top}\in R^{p}$, with
all $\xi_{i}>0$, $i=1,\cdots,p$, and
$\xi_{0}=[\xi_{p+1},\xi_{p+2},\cdots,\xi_{m}]^{\top}\in R^{m-p}$
with all $\xi_{j}=0$, $p+1\le j\le m$. Then, $A$ can be rewritten
as $\left[\begin{array}{cc}A_{11}& A_{12}\\A_{21}&
A_{22}\end{array}\right]$  where $A_{11}\in R^{p,p}$ is
irreducible and $A_{12}=0$.
\end{enumerate}
For detail proof, readers can refer to [5].

{\bf Remark 4} Adaptive algorithm in Theorem 2 of [2] is identical to that reported in Theorem 5 in the section "Adaptive adjustment of the coupling strength" of [1].

\begin{eqnarray}
\left\{\begin{array}{cc}\frac{dx_1(t)}{dt}&=f(x_1(t),t)
+c(t)\sum\limits_{j=1}^ma_{1j}x_j(t)\\
&-c(t)\varepsilon(x_{1}(t)-s(t)),\\
\frac{dx_i(t)}{dt}&=f(x_i(t),t)+c(t)\sum\limits_{j=1}^ma_{ij}x_j(t),\\
&i=2,\cdots,m\\
 &\dot{c}(t) =\frac{\alpha}{2}\sum\limits_{i=1}^m\delta
x_i^T(t)P\delta x_i(t)\end{array}\right. 
\end{eqnarray}
where $c(0)\geq 0$ and $\alpha>0$, can synchronize the coupled
system to the given trajectory $s(t)$.

{\bf Remark 5} It is clear that Lyapunov function in Theorem 1 of [2]
\begin{align}
V(t)=\sum_{i=1}^{m}e_{i}(t)^{T}e_{i}(t)
\end{align}
is a special case of the Lyapunov function
\begin{align}
V(t)=\sum_{i=1}^{m}\delta x_{i}(t)^{T}P\delta x_{i}(t)
\end{align}
reported in Theorem 2 of [1].

{\bf Remark 6} It is clear that Lyapunov function in Theorem 3.1 of [3]
\begin{align}
V(t)=\sum_{i=1}^{m}\xi_{i}e_{i}(t)^{T}e_{i}(t)
\end{align}
is a special case of the Lyapunov function
\begin{align}
V(t)=\sum_{i=1}^{m}\xi_{i}\delta x_{i}(t)^{T}P\delta x_{i}(t)
\end{align}
reported in Theorem 3 in [1].

{\bf Remark 7}
The section 4 in [3] "Synchronization criteria via pinning control on networks with a directed spanning tree" has been reported in [1] in detail.

See section "C. Pin A Linearly Coupled Network With A Reducible
Asymmetric Coupling Matrix" in section II of [1].

In the following, we remove the assumption that $A$
is irreducible. In this case, we assume
\begin{eqnarray}
A=\left[\begin{array}{cccc}A_{11}&0&\cdots&0\\
A_{21}&A_{22}&\cdots&0\\
\vdots&\ddots&\vdots&\vdots\\
A_{p1}&A_{p2}&\cdots&A_{pp}\end{array}\right] 
\end{eqnarray}
where   $A_{qq}\in R^{m_{q},m_{q}}$, $q=1,\cdots,p$,  are
irreducible. And, for each $q$, there exists $q>k$ such that
$A_{qk}\ne 0$. It is equivalent to that the connecting graph has a
spanning tree.

It is easy to see that if we add a single controller
$-\epsilon(x_{1}(t)-s(t)$ to the node $x_{1}(t)$. Then, by
previous arguments, we conclude that the subsystem
\begin{eqnarray}
\left\{\begin{array}{cc}\frac{dx_1(t)}{dt}&=f(x_1(t),t)+c\sum\limits_{j=1}^{m_{1}}a_{1j}x_j(t)\\
&-c\varepsilon(x_{1}(t)-s(t)),\\
\frac{dx_i(t)}{dt}&=f(x_i(t),t)+c\sum\limits_{j=1}^{m_{1}}a_{ij}x_j(t),\\
&i=2,\cdots,{m_{1}} \end{array}\right.
\end{eqnarray}
pins $x_{1}(t),\cdots,x_{m_{1}}$ to $s(t)$.

Now, for the subsystem $x_{m_{1}+1},\cdots,x_{m_{2}},$ we have
\begin{eqnarray*}
&&\frac{d\delta
x_i(t)}{dt}=f(x_i(t),t)-f(s(t),t)+c\sum\limits_{j=1}^{m_{2}}a_{ij}\delta
x_j(t)
\\&=&f(x_i(t),t)-f(s(t),t)+c\sum\limits_{j=m_{1}+1}^{m_{2}}a_{ij}\delta x_j(t)+O(e^{-\eta
t})
\end{eqnarray*}

Because $A_{21}\ne 0$. Then, in $A_{22}$, there exists at least
one row $i_{2}$, such that
\begin{eqnarray}
a_{i_{2}i_{2}}>\sum_{j=m_{1}+1}^{m_{2}}a_{i_{2}j}
\end{eqnarray}
Therefore, all eigenvalues of $A_{22}$ are negative. By the
similar arguments in the proof of theorem 4, we can pin
$x_{i}(t)$, $i=m_{1}+1,\cdots,m_{2}$, to $s(t)$.

By induction, we prove that if we add a controller to the master
subsystem corresponding to the sub-matrix $A_{11}$, then we can
ping the complex network to $s(t)$ even if the coupling matrix is
reducible.

{\bf Conclusions}
In this note, we point out that most of the results in papers [2] and [3] have been reported and are direct consequences of previous papers [1], [4]. We also correct the proof given in the papers [2] and [3].
\newline

\noindent{\bf References}

[1].
Chen, T., Liu, X.,  and Lu,W., "Pinning Complex Networks by a Single Controller", IEEE Transactions on Circuits and Systems-I: Regular Papers, 54(6), 1317-1326  (2007)

[2]. Yu.,W., Chen,G., Lu.,J., "On pinning synchronization of complex dynamical networks", Automatica 45  429-435 (2009)

[3]. Yu.,W., Chen,G.,Lu.,J., and Jurgen Kurth, "Synchronization via pinning control on general complex networks", SIAM J. CONTROL OPTIM. 51(2), 1395-1416 (2013)

[4]. Wenlian Lu, Tianping Chen, "New Approach to Synchronization
Analysis of Linearly Coupled Ordinary Differential Systems", Physica D, 213, 214-230  (2006)

[5]. Wenlian Lu and Tianping Chen, "A New Approach to Synchronization Analysis of Linearly Coupled Map Lattices", Chinese Annals of Mathematics, 28B(2), 149-160 (2007) 

\end{document}